\documentclass{ws-jtca}
\usepackage[super]{cite}

\usepackage{xcolor}

\newcommand{\vb}[1]{\mathbf{#1}} 
\newcommand{\be}{\begin{equation}}
\newcommand{\ee}{\end{equation}}
\newcommand{\ben}{\begin{equation*}}
\newcommand{\een}{\end{equation*}}
\newcommand{\TS}{\text{TS}}
\newcommand{\sinc}{\mathrm{sinc}}

\begin{document}
\markboth{Cascallares, Gonzalez \& Lavia}{TetraScatt model}
%
\catchline{}{}{}{}{}
%
\title{TetraScatt model:
 Born approximation for the estimation of acoustic dispersion of fluid-like objects of arbitrary geometries.}
\author{Edmundo F. Lavia$^{(1,2)}$, Guadalupe Cascallares$^{(3,4)}$, Juan D. Gonzalez$^{(1,2)}$}
\address{
$^1$\text{Acoustic Propagation Department. Argentinian Navy Research Office (DIIV)}\\
\text{Laprida 555 Vicente Lopez, (1638) Buenos Aires, Argentina}\\
$^2$\text{UNIDEF (National Council of Scientific and Technical Research – Ministry of Defense)}\\
$^3$\text{National Institute for Fisheries Research and Development (INIDEP)}
\\ \text{Paseo Victoria Ocampo 1, Escollera Norte, Mar del Plata, Argentina.}\\
$^4$\text{CONICET}.
}

\maketitle

\begin{history}
\received{(10 May 2022)}
\today
\end{history}

\begin{abstract}Modelling the acoustic scattering response due to penetrable objects of arbitrary shapes, such as those of many marine organisms, can be computationally intensive, often requiring high-performance computing equipment when considering a completely general situation. However, when the physical properties (sound speed and density) of the scatterer object under consideration are similar to those of the surrounding medium, the Born approximation provides a computationally efficient way to calculate the scattering.
For simple geometrical shapes like spheres and spheroids, the acoustic scattering in the far field can be evaluated through the Born approximation recipe as a formula which has been historically employed to predict the response of weakly scattering organisms, such as zooplankton. Moreover, the Born approximation has been extended to bodies whose geometry can be described as a collection of non-circular rings centred on a smooth curve.
In this work, we have developed a numerical approach to calculate the far-field backscattering by arbitrary 3D objects under the Born approximation. The object's geometry is represented by a volumetric mesh composed of tetrahedrons, and the computation is efficiently performed through analytical 3D integration, yielding a solution expressed in terms of elementary functions for each tetrahedron. The method's correctness has been successfully validated against benchmark solutions. Additionally, we present acoustic scattering results for species with complex shapes.
To enable other researchers to use and validate the method a computational package named \texttt{tetrascatt} implemented in the R programming language was developed and published in the CRAN (Comprehensive R Archive Network).
\end{abstract}


\section{Introduction}

Acoustic sampling has proven to be a useful tool to estimate biomass in fishery acoustics.
The physical and mathematical theory on which it is based was established by the pioneering works of some of most important researchers from XIX century and consequently methods and techniques aimed to solve the acoustic scattering problem when certain simplifying assumptions are fulfilled are bonded to such {\it big} names as Helmholtz, Kirchhoff or Born.

When the shape of the body which produces the scattering can be approximated by a simple geometrical shape (sphere, spheroid or cylinder, for instance) the previously mentioned approaches even provide an analytical solution.
However, the bodies of interest in fisheries acoustics are usually not simple geometries, so it is necessary fall to numerical techniques.

Accurate computation of scattering from an arbitrary shaped obstacle can be a challenging task, as it requires solving the exterior Helmholtz equation in three dimensions under appropriate boundary conditions. When the scattering body can be considered as an impenetrable one, a reasonable approximation may be the Kirchhoff approximation \cite{lavia2018modelling}. It gives valid solutions for (mainly) convex bodies in the high frequency regime, and its application avoids solving many of the problems that occur when calculating the exact solution, such as the resolution of high-dimensional matrix systems and the computation of singular integrals.

When the scattering objects are penetrable, but their physical properties do not vary significantly with respect to the properties of the medium where the wave propagates, an approach known as the Born Approximation (BA) can be made \cite{morse1986theoretical}.
The conditions of this approximation are what is known in the literature as {\it weakly scattering}. The key assumption is that, up to first order, the perturbation induced for the impinging wave in the interior of this weakly scattering object is neglected. As a result, the acoustic field within the object is assumed to be identical to the incident field. This allows us to solve the scattering through an explicit evaluation of an integral, without the need to solve an integral equation or its discretized version.

Improved versions of the original BA are the Distorted Wave Born Approximation (DWBA), and also its phase compensated version \cite{chu1999phase,jones2009use} which are of widespread use in many works regarding scattering by weakly organisms as zooplankton.
The first models focused on marine organisms were based on canonical shapes \cite{stanton1993average}.
Then different schemes allows to deal with complex bodies in an approximated way, see \cite{stanton2000review,jones2009use} and references therein.
Further, use of the Born approximation was extended to bodies whose geometry can be described as a collection of non-circular rings centred on a smooth curve \cite{gastauer2019zooscatr}. Such approaches are not applicable to incidences far from the broadside incidence. 

A work addressing the BA for arbitrary shaped geometries is presented in the Ref. \citen{pees2011efficient} where the pressure field is calculated resorting to a fundamental theorem on diffraction tomography which involves Fourier transforms.

In this work, a numerical approach to calculate the far field backscattering by arbitrary three-dimensional objects under the Born approximation is developed and computationally implemented in the R programming language.
The model is based on direct (analytic) integration over a discretized version of the scattering volume, which only requires elementary techniques avoiding thus the use of quadrature rules. In this way a fast and general approach is obtained. A computational code is made publicly available as an R package named \texttt{tetrascatt}\cite{paquete} in the CRAN (Comprehensive R Archive Network) repository.
The TetraScatt model is validated using as benchmarks exact solutions of the Born approximation and exact solutions of general scattering problems evaluated in weakly conditions. Then the code is used to explore the validity range of the Born approximation itself.

This work is organized as follows, in the Section \ref{model} a brief summary of the BA is provided followed by a description of the mathematical and numerical implementation of the model.
Section \ref{validation} shows a set of tests against benchmark solutions which gives a validation for the model.
In Section \ref{validity} numerical experiments are conducted to explore the range of validity where the BA gives predictions with an acceptable error margin. Exact solutions of general scattering are employed for this task.
As an application of the model, a backscattering calculation from a native Argentinian species of crustacean is presented in Section \ref{peisos}.
Finally, conclusions of the work are drawn in Section \ref{conclusions}.

\section{Model description}
\label{model}

\subsection{Scattering by the Born approximation}

\begin{figure}[!ht]
	\centering
	\includegraphics[scale=0.5]{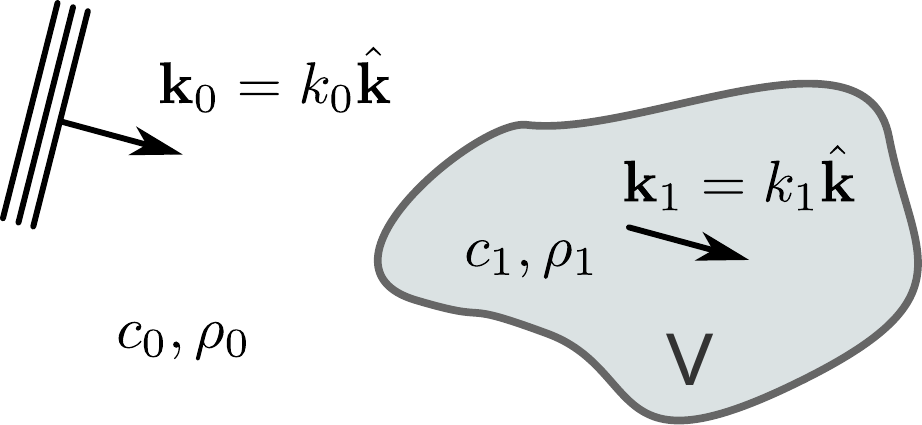}
	\caption{Geometry of the scattering by an arbitrary body of volume $V$.}
	\label{fig_esquema_born}
\end{figure}

The acoustic scattering problem we considered here is depicted in Fig. \ref{fig_esquema_born}. A plane harmonic wave of frequency $f$ propagating in a medium of sound speed and density $c_0,\rho_0$ is incident on a scattering object with internal properties $c_1,\rho_1$.
Two wavenumbers $k_i=2\pi f/c_i$ (with $i=0,1$) are thus defined.

The backscattering far-field form function $f_\infty$ under the Born approximation\cite{morse1986theoretical} can be expressed as the volume integral
\begin{equation}
	f_\infty^\text{BA} = \frac{k_0^2}{4\pi} \: \int_V \: ( \gamma_\kappa - \gamma_\rho ) \: e^{2 i k_0\hat{\bf{k}} \cdot \bf{x}} \: dV(\bf{x}),
	\label{ba_finf}
\end{equation}
where the factors $\gamma$ are related to the density $\rho$ and compressibility $\kappa = 1 / (\rho c)^2$ of the two media and can be expressed in terms of the
ratios $g \equiv \rho_1/\rho_0$ and $h \equiv c_1/c_0$, according to
\begin{equation*}
	\gamma_\rho = \frac{\rho_1 - \rho_0}{\rho_1} = \frac{g-1}{g} \qquad \qquad
	\gamma_\kappa = \frac{\kappa_1 - \kappa_0}{\kappa_0} = \frac{1-gh^2}{gh^2}.
\end{equation*}

Since the Born approximation assumes similar properties for the exterior media ``0'' and interior media ``1'', it follows that $g \approx 1$ and $ h \approx 1$.
A modified version of this approach is the distorted wave Born approximation (DWBA)\cite{chu1999phase} which replaces the wave vector $\mathbf{k}_0$ in both the phase of the internal field and the Green function\footnote{After performing the far-field limit.} by $\mathbf{k}_1$, thus representing a {\it distorted wave}.

This modification provides, in the backscattering calculations, a slightly improvement over the original expression.
For homogeneous media the DWBA results in
\begin{equation}
	f_\infty^\text{DWBA} = \frac{k_1^2}{4\pi} \: ( h^2 \gamma_\kappa - \gamma_\rho ) \:
			\int_V \: e^{2 i k_1 \hat{\bf{k}} \cdot \bf{x}} \: dV(\bf{x}).
	\label{dwba_finf}
\end{equation}

For completeness, it is worth mentioning that exists at least one alternative realization\cite{jones2009use} of the DWBA which in the far-field expression carry out the change $k_0 \to k_1$ in all the places where it appears. This alternative expression differs from Eq. \eqref{dwba_finf} in the absence of the $h^2$ factor affecting the $\gamma_\rho$ term.

\subsection{TetraScatt formulation}

The TetraScatt model calculates the DWBA backscattering due a fluid volume $V$ by discretizing it in a collection $ V = \cup_{j=1}^N \: T_j$ of $N$ tetrahedra $T_j$ and then evaluating the integration defined in \eqref{dwba_finf} as the sum of integrals over each tetrahedron.
Then, the integral of Eq. \eqref{dwba_finf} results in
\begin{equation}
	f_\infty = \sum_{j=1}^N \:
		\frac{k_1^2}{4\pi} \: ( h^2\gamma_\kappa - \gamma_\rho ) \: \int_{T_j} \: e^{2 i k_1 \hat{\bf{k}} \cdot \bf{x}} \: dV(\bf{x}),
	\label{dwba_finf_discreto}
\end{equation}
where $\bf{x}$ are the coordinates of the $T_j$ tetrahedron.

\begin{figure}[!thb]
	\begin{center}
	\includegraphics[width=0.25\textwidth]{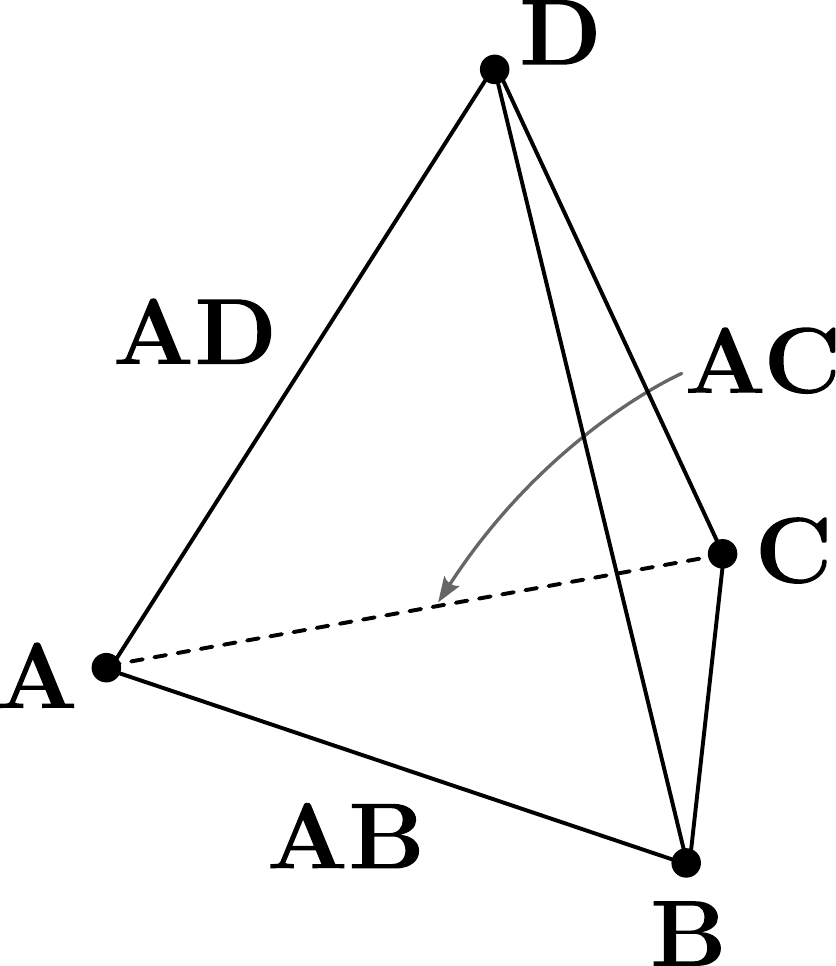}
	\end{center}
	\caption{Arbitrary tetrahedron of vertex $\vb{A}, \vb{B}, \vb{C}, \vb{D}$ which define, with respect to vertex $\vb{A}$, the oriented segments $\vb{AB}, \vb{AC}, \vb{AD}$.}
	\label{fig_tetrahedro_ref}
\end{figure}

The arbitrary chosen vertex $\vb{A},\vb{B},\vb{C},\vb{D}$ of each tetrahedron (see Figure \ref{fig_tetrahedro_ref}, where a generic tetrahedron is illustrated) define oriented segments $\vb{AB} = \vb{B}-\vb{A}, \vb{AC} = \vb{C}-\vb{A}, \vb{AD}=\vb{D}-\vb{A}$ which can be used for its parameterization.

Then, integration over the $j$-th tetrahedron with vertex $\vb{A}_j,\vb{B}_j,\vb{C}_j,\vb{D}_j$ is carried out using the parameterization
\be
	\vb{x}: R \to T_j \subseteq \mathbb{R}^3, \quad
	\vb{x}(u,v,w) = \vb{A}_j + w \: \vb{AB}_j + v \: \vb{AC}_j + u\vb{AD}_j,
	\label{param_T}
\ee
where $R\subset \mathbb{R}^3$ is defined as  
\be
	R = \{(u,v,w): 0 \leq u + v + w \leq 1 \quad \text{with} \quad u,v,w \geq 0 \},
\ee
which describes the j-th tetrahedron referred to the vertex $\vb{A}$, arbitrarily chosen as origin of the local coordinates.

The integration over a single tetrahedron appearing in Eq. \eqref{dwba_finf_discreto} in terms of the parametrization \eqref{param_T} reads
\be
	\int\limits_{T_j} \: e^{2 i k_1\hat{\vb{k}} \cdot \vb{x}} \: dV(\vb{x}) = J
	\int\limits_{R } 
	e^{2 i k_1\hat{\vb{k}} \cdot (\vb{A}_j + w\vb{AB}_j + v \vb{AC}_j + u \vb{AD}_j) }dV = J  F(\vb{A}_j,\vb{B}_j,\vb{C}_j, \vb{D}_j,\hat{ \bf{k}},k_1)
	\label{integral_born_app}
\ee
where $J=J(\vb{A}_j,\vb{B}_j,\vb{C}_j, \vb{D}_j)$ is the Jacobian accompanying the volume differential, which is constant within the tetrahedron $T_j$, and $F$ is an elementary function based on polynomials an exponential functions only, whose detailed expression is given in the Appendix \ref{app_F}.

The discretization of a volume $V$ in tetrahedra can be carried out in a CAD software as SALOME\cite{ribes2007salome} which will result in a volumetric mesh (a structure containing information about the vertex, the tetrahedra which constitutes the volume and the plane triangles which build the surface of the enclosed volume).

Because a tetrahedron has sides which are plane triangles any curvature in the scattering body surface has to be represented in an approximated way through a discretization that includes enough elements to ensure that curvature is captured by them.

In summary, the TetraScatt model takes as input: a mesh representing the geometry of the scattering body, the physical parameters of the involved media and the scattering parameters.

\section{Model validation}
\label{validation}

We perform two types of verification for the TetraScatt model:
(a) To ensure the accuracy of the integral computation implementation, we compare the model's output against exact DWBA integral solutions. This comparison is conducted for cases where the DWBA integral expressed in Eq. \eqref{dwba_finf} has a closed-form expression, which happens for example when the scatterer's geometry is an sphere or an spheroid.
(b) In order to assess the validity of the presented model as a solution of the acoustic scattering problem for arbitrary shaped objects we compare our results with a well established benchmark solutions for penetrable acoustic scattering \cite{jech_comparisons_2015}. Specifically, we utilize the Boundary Element Method (BEM) solution \cite{gonzalez2020boundary} and the method based on spheroidal wave functions \cite{gonzalez2016computational} which can solve the scattering by prolate and oblate spheroids. While the BEM solution is not constrained by material properties or geometrical shapes of the scatterers, making it applicable beyond weak scattering scenarios, it is both computationally and theoretically more intricate.
As a test of rigorous scattering modelling without any assumptions we need to carry on these tests in the $g, h$ domain where the Born prescription gives a negligible error, so that any departure can be attributed to the TetraScatt formulation and not to a breakdown of the Born approach itself.

The first two subsections (\ref{val_sphere} and \ref{val_prolate}) deal with analytical DWBA solutions (the (a)-type verifications), while subsection \ref{val_copepodo} presents a comparison with BEM results for weakly scattering by a copepod (a (b)-type verification).
For all cases, we chose media parameters with values used in Ref. \citen{jech_comparisons_2015}, as listed in the Table 1. The surrounding media properties are those of seawater.

\begin{table}[th]
\label{tabla_propiedades1}
	\begin{center}
		\begin{tabular}{@{}ccccc@{}} \toprule
		Parameter & $c$ (m/s) & $\rho$ (g/cm$^3$) &
		$c_1$ (m/s) & $\rho_1$ (g/cm$^3$) \\ \colrule
		Value & 1477.4\hphantom{0} & 1026.8\hphantom{0} & 1480.3\hphantom{0} & 1028.9\hphantom{0}\\ \botrule
		\end{tabular}
	\end{center}
	\caption{Material properties (density $\rho$ and sound speed $c$) of the targets and surrounding medium.}
\end{table}

A comment on the nature of the exact benchmark solutions used in this work is relevant here. The far-field scattering by a fluid sphere, the classical partial-wave solution of Ref. \citen{anderson_sound_1950}, which also can be found in Ref. \citen{jech_comparisons_2015}, is constituted by a modal series solution whose coefficients has an explicit form given in terms of spherical Bessel functions.
The exact far-field solution for prolate and oblate spheroids is far more involved than the spherical one since the computation of the spheroidal wave functions is a more difficult task than the Bessel function computation.
Furthermore, the expansion coefficients has not analytic expression and are to be founded by solving a family of recursive matrix systems.
Bessel function evaluation is integrated in almost any computational environment aimed to scientific numerical usage. On the contrary, spheroidal function evaluation is restricted to a few, specific, implementations. In this work we adopt the model proposed in Ref. \citen{lavia2021acoustic}, which is based on arbitrary-precision arithmetic for computing the modal series of prolate spheroidal wave functions, aimed to a parallel computation environment of high performance computing.
The BEM requires a mesh representation of the scatterer surface, usually in the form of a collection of $N$ planar triangles. The solution of the fluid scattering problem case results in a matrix system whose size is $4N^2$. Given that the length $\ell$ of the segments which constitutes each triangle in this mesh is required to fulfill a condition $\ell \leq \lambda/5$, where $\lambda$ is the minimum implied wavelength ($\lambda = \mathrm{Min}\left\{ \lambda_0=c_0/f,\lambda_1=c_1/f \right\})$), the size $N$ increases with an increase in the frequency $f$.

\subsection{Sphere}
\label{val_sphere}

Computing DWBA for a sphere of radius $a$ through Equation (\ref{dwba_finf}) gives\cite{chu1999phase}
\begin{equation}
	f_\infty = k_1^2 a^3 \: ( h^2\gamma_\kappa - \gamma_\rho ) \: \frac{j_1(2k_1a)}{2k_1a},
	\label{dwba_finf_sphere}
\end{equation}
being $j_1$ the spherical Bessel function of first order.

\begin{figure}[!th]
	\centering
     \includegraphics[scale=0.18]{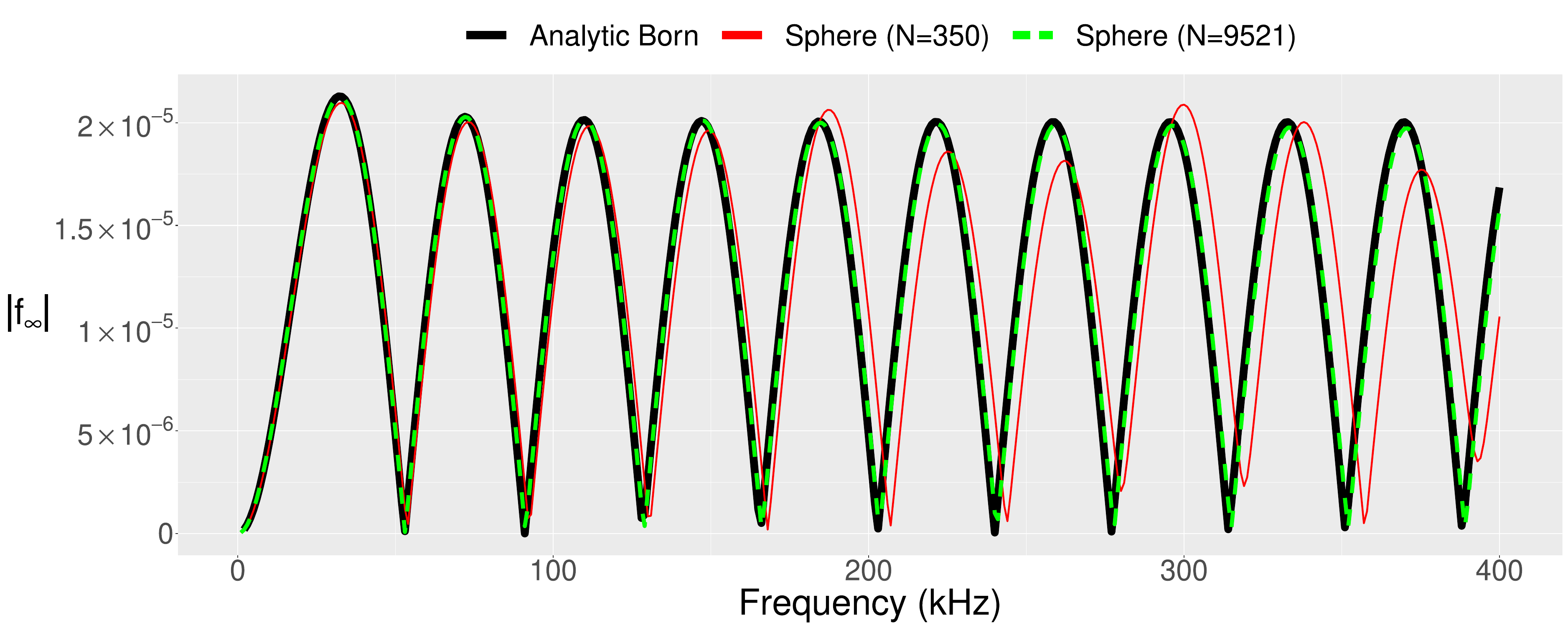}
	\caption{Absolute value $|f_\infty|$ for the backscattering by an sphere under the DWBA according to the exact formula \eqref{dwba_finf_sphere} and two executions of TetraScatt for meshes with the indicated number $N$ of tetrahedra.}
	\label{fig_validacion_esferas}
\end{figure}

For the comparison with the TetraScatt model, two meshes with $N=350$ and $N=9521$ tetrahedra representing a sphere of $a = 0.01$ m have been used. Two tests has been conducted. In first place an evaluation of the backscattering $|f_\infty|$ for the frequency range $(1,400)$ kHz was carried out. The three resulting curves are shown in Fig. \ref{fig_validacion_esferas}.
The coarse mesh $N=350$ shows increasing departure from the exact solution of Eq. \eqref{dwba_finf_sphere} whereas the finer one is in good agreement along the entire frequency range. As expected, as long as the number of tetrahedron in the mesh increases the error respect the exact solution decreases.

\begin{figure}[!hb]
	\centering
	\includegraphics[scale=0.20]{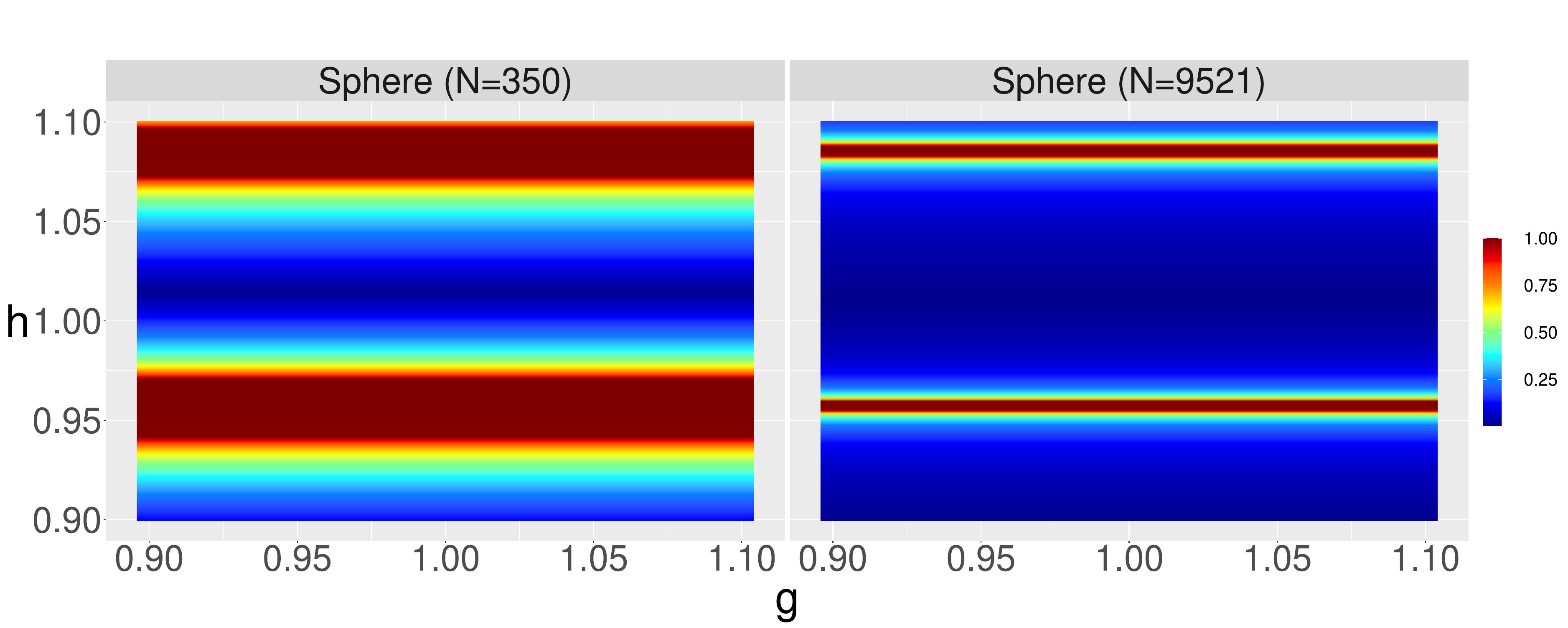}
	\\\vspace{2mm}
	\includegraphics[scale=0.17]{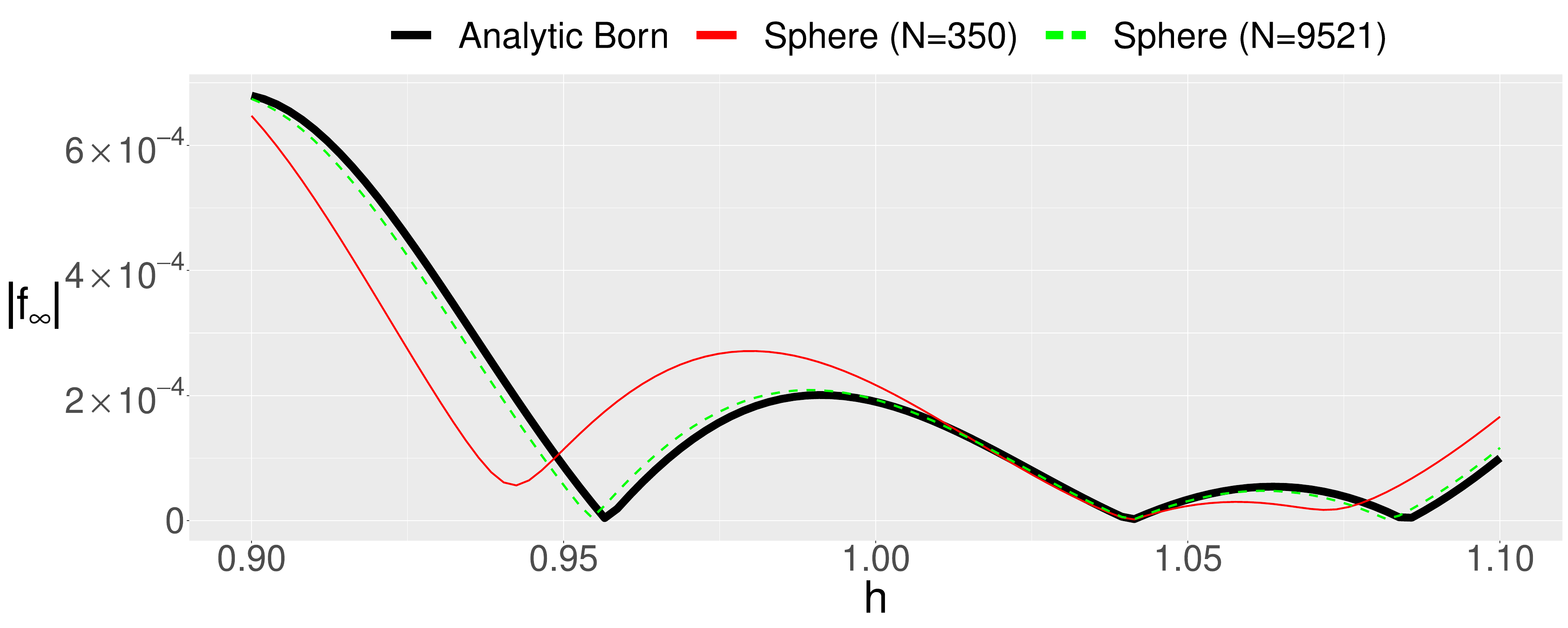}
	\caption{Relative error $|f_\infty-f_\infty^\text{ex}| / |f_\infty^\text{ex}|$ for the backscattering with TetraScatt with respect to the exact formula \eqref{dwba_finf_sphere} executed on the $N=350$ mesh (top left) and on the $N=9521$ mesh (top right) at $f = 300$ kHz.
	Both plots were saturated at 1 to provide a comparison on equal footing.
	The two error peaks seen at $h = 0.956$ and $h = 1.083$ are related to nulls on the acoustic response as can be verified by the plot (bottom) at $g=0.96$. The point $g=h=1$ (zero error by definition there) has been removed from both top plots.}
	\label{fig_error_born_heatmap}
\end{figure}

In second place, an evaluation of the relative error $E = |f_\infty-f_\infty^\text{ex}| / |f_\infty^\text{ex}|$ for the range $0.9 \leq g \leq 1.1$ and $0.9 \leq h \leq 1.1$ at fixed frequency $f=$ 300 kHz was carried out, being the results exhibited in the plots at top of the Fig. \ref{fig_error_born_heatmap} as heatmaps.
The colorbar scale was saturated at 1 in both plots to facilitate a direct  comparison. The behaviour of error is evidently influenced by $h$ as indicated by the  translational symmetry observed in both plots along the $g$-direction.
There are two critical values $h=0.956$ and $h=1.083$ where local maxima is attained for all $g$-values. They are associated to nulls offset as depicted in the plot at the bottom of the figure.

\subsection{Prolate spheroid}
\label{val_prolate}

\begin{figure}[!bh]
	\centering
	\includegraphics[scale=0.18]{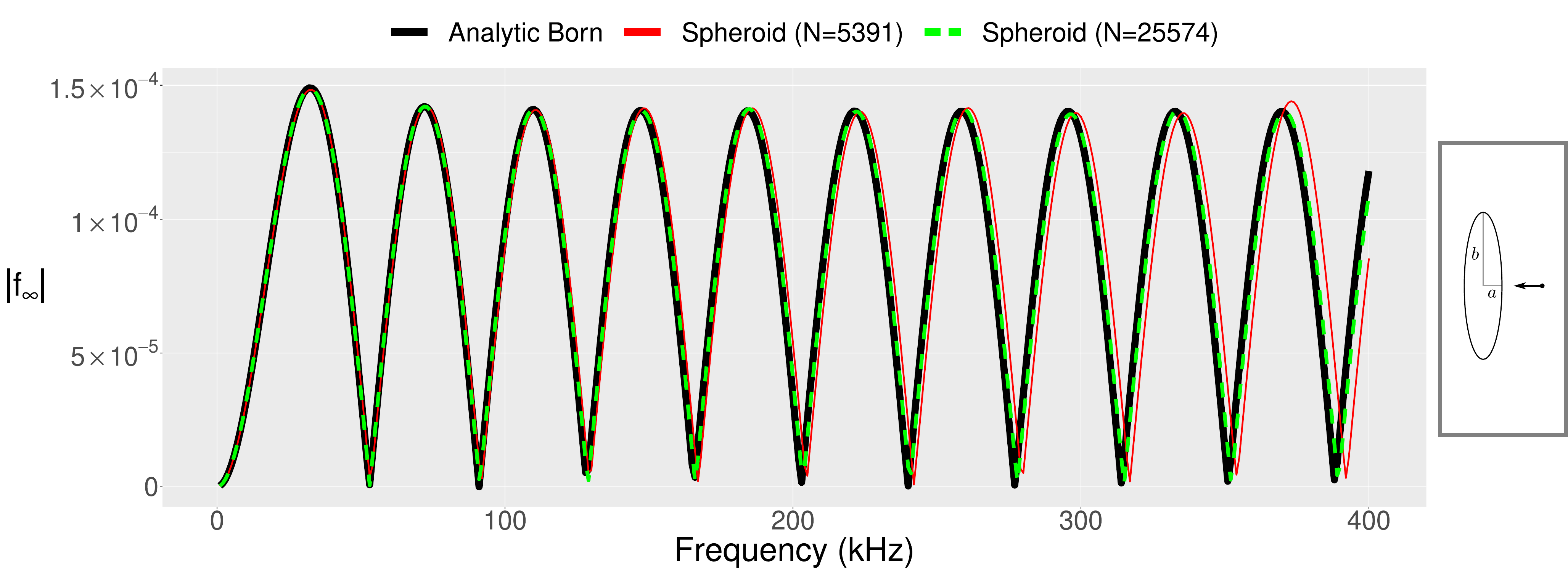}
	\includegraphics[scale=0.18]{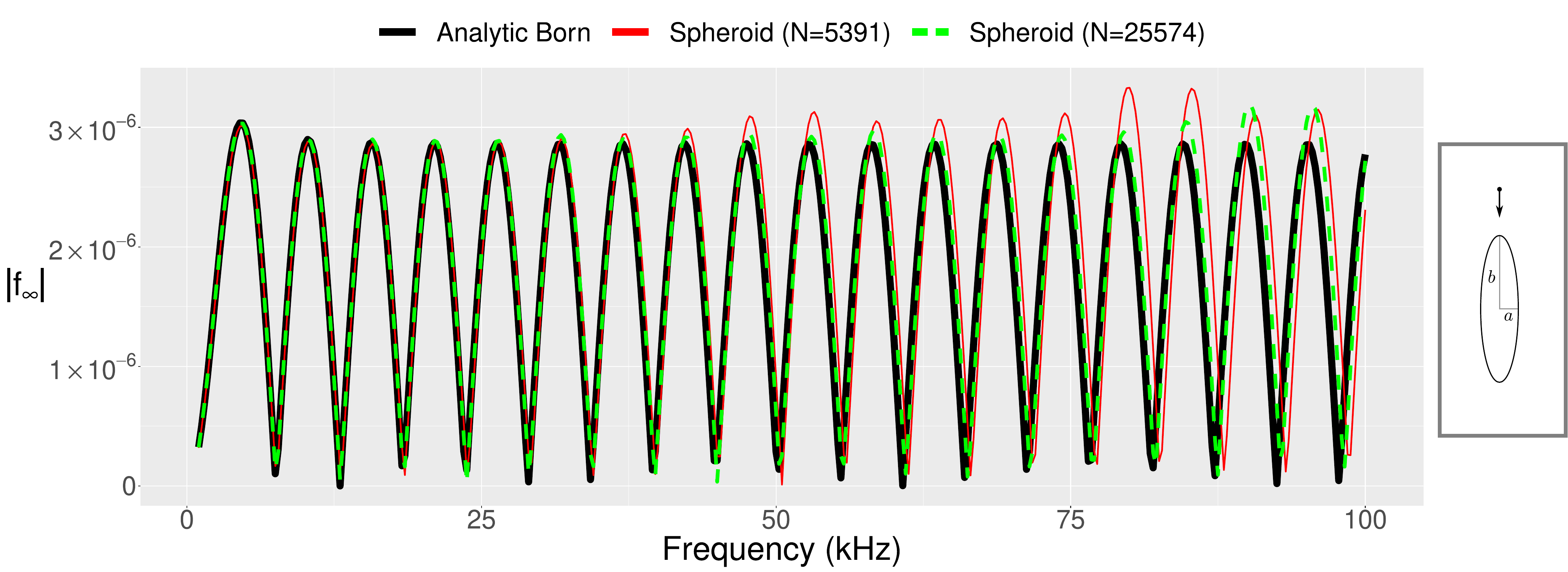}
	\caption{Absolute value $|f_\infty|$ for the backscattering by an spheroid under the DWBA according to the exact formula \eqref{dwba_finf_spheroid} and two executions of TetraScatt with the indicated meshes. Top: broadside incidence. Bottom: end-on incidence.}
	\label{fig_validacion_esferoides}
\end{figure}

For a prolate spheroid \cite{chu1999phase} of minor and major semi-axis $a,b$, respectively, the DWBA has an especially simple expression in the mutually perpendicular cases of broadside and end-on incidences. In these cases it results in
\begin{equation}
	f_\infty^\text{DWBA} =
			\begin{cases}
			\displaystyle k_1^2 a^3 e \: ( h^2\gamma_\kappa - \gamma_\rho ) \:
					\frac{j_1(2k_1a)}{2k_1a}, \qquad \text{Broadside incidence}\\
			\\
			\displaystyle k_1^2 a^3 e \: ( h^2\gamma_\kappa - \gamma_\rho ) \:
					\frac{j_1(2k_1b)}{2k_1b}, \qquad \text{End-on incidence}
			\end{cases}
	\label{dwba_finf_spheroid}
\end{equation}
where $e=b/a>1$ is the spheroid aspect-ratio. The sphere solution is recovered when $e\to 1$ (see Eq. \eqref{dwba_finf_sphere}).
For the comparison against the model two meshes representing an spheroid with $a=0.01$ m and $b=0.07$ m of $N=5391$ and $N = 25574$ tetrahedra have been used.
The Fig. \ref{fig_validacion_esferoides} shows evaluation of the backscattering $|f_\infty|$ for the ranges $(1,400)$ and $(1,100)$ kHz in broadside and end-on incidences, respectively (inserts at the right of each figure illustrates the incidence direction).

The results exhibit qualitatively similar behaviour to the sphere case: an error that increases with frequency, and finer meshes providing better agreement than coarser ones. When compared with the sphere, the spheroid requires meshes with a higher number of tetrahedra for \texttt{TetraScatt}, attributed to its larger surface area and the strong curvature imposed by the high aspect ratio $e=7$ (refered as 1:7).

\subsection{Weakly scattering by a copepod}
\label{val_copepodo}

The two examples previously used as a validation of the model show that if the number of tetrahedral elements in the volumetric mesh is high enough to ensure a proper external curvature representation the error can be negligible.

In order to evaluate the TetrasScatt model with a complex scattering object, a profile representative of a copepod available in the shape catalog inside ZooScatR package \cite{gastauer2019zooscatr} was used. Using this profile as a curved-axis of symmetry a scattering body can be defined.
Since there is not analytical solution for this kind of geometry, we use the BEM implementation of Ref. \citen{gonzalez2020boundary} as a benchmark.

\begin{figure}[!ht]
 	\centering
	\includegraphics[scale=0.20]{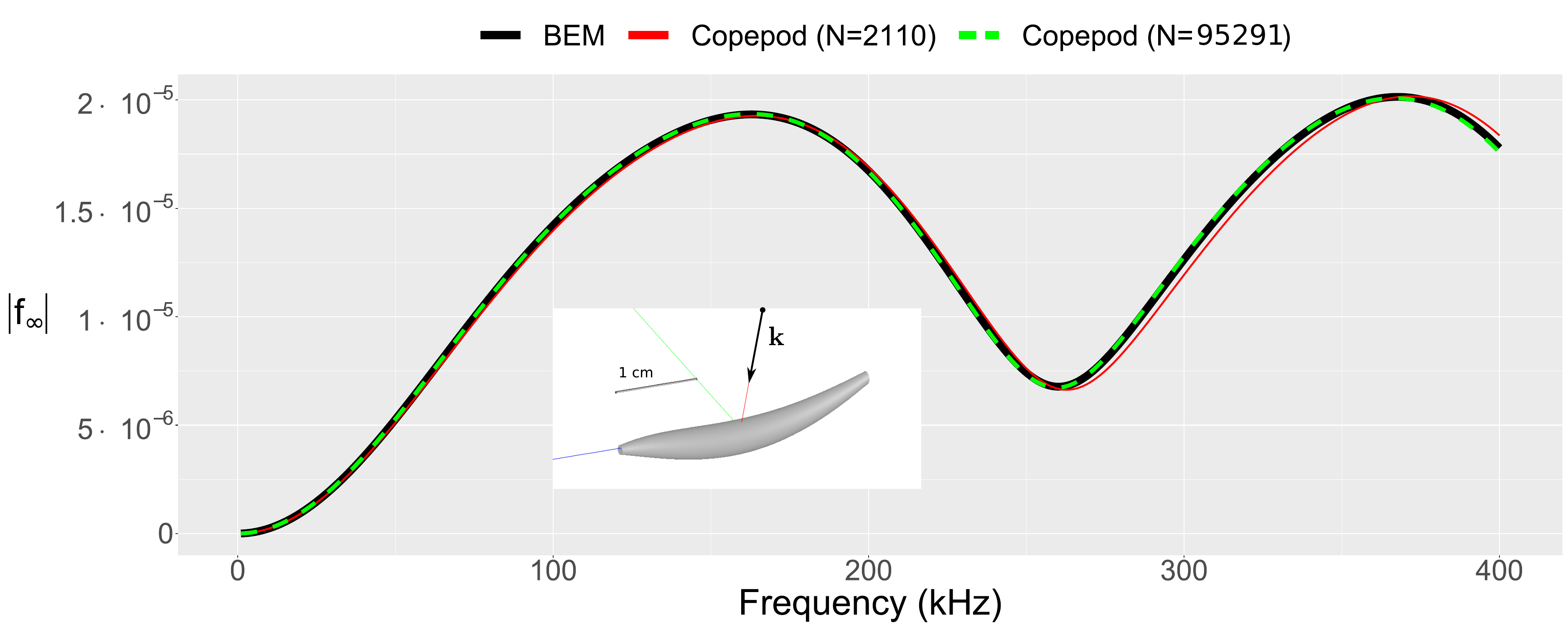}
	\caption{Absolute value $|f_\infty|$ for the backscattering by a idealized marine organism: a copepod.
	The BEM solution was taken as the benchmark whereas for the TetraScatt execution two meshes of the indicated tetrahedral elementre were used.
	The insert in the main window shows the mesh, the reference system and the incidence direction.}
	\label{fig_validacion_copepodo}
\end{figure}

For the TetrasScatt model two meshes of $N=2110$ and $N=95291$ were drawn from the geometry shown in the insert in the plot on Fig. \ref{fig_validacion_copepodo} where also are indicated the incidence direction and a reference scale. Absolute value $|f_\infty|$ as a function of frequency from 1 to 400 kHz is showed in the main window of the figure for the model and the BEM benchmark.
The physical parameters used were the indicated in Table 1. For the BEM evaluation a mesh with $N = 6608$ triangular elements was enough to evaluate the entire frequency interval.

\section{Validity range of the DWBA for $g$ and $h$}
\label{validity}

There is general agreement between researchers about the validity of the BA and the DWBA; when the contrast parameters
$g, h$ differ even slightly from unity the error increases significantly.
Further, this behaviour is uneven in $g$ and $h$ and non-monotonous.

In fisheries acoustics, as well as in other marine applications, sound scattering by an object is analyzed in the logarithmic scale using the target strength (TS) parameter which can be expressed as
\be
	\TS = 10 \log_{10} \left( |f_\infty|^2 \right) \text{ dB re 1 m$^2$}.
	\label{TS_def}
\ee
In addition, since sea measurements are affected by a large number of factors upon which researchers have little control, TS scattering predictions with a certain amount of error are allowed. In this vein a more appropriate, and permissive, quantification for the model error can be defined\cite{jech_comparisons_2015} as
\begin{equation}
	\varepsilon = \left| \TS^\text{model} - \TS^\text{benchmark} \right|.
	\label{error_ts}
\end{equation}

In the next sub-sections predictions of the TetraScatt model for the range $0.9 \leq g \leq 1.1$ and $0.9 \leq h \leq 1.1$ will be compared with appropriate benchmark solutions and errors according to the Eq. \eqref{error_ts}.
Two benchmark solutions of the general, i.e. not weakly, fluid acoustic scattering problem were used: the farfield scattering by a fluid sphere, the classical partial-wave solution\cite{anderson_sound_1950,jech_comparisons_2015} and the scattering by a prolate fluid spheroid\cite{gonzalez2016computational} under the end-on incidence.

The sphere and spheroid keep the given values at sections \ref{val_sphere} and \ref{val_prolate}, respectively.
To ensure a proper curvature representation in both the sphere and spheroid, meshes of an appropriate number of tetrahedral elements were used. This is crucial in the spheroid case since end-on incidence is the computationally hardest direction when high aspect ratios are considered.


\subsection{Sphere}

\begin{figure}[!hb]
	\centering
	\includegraphics[scale=0.30]{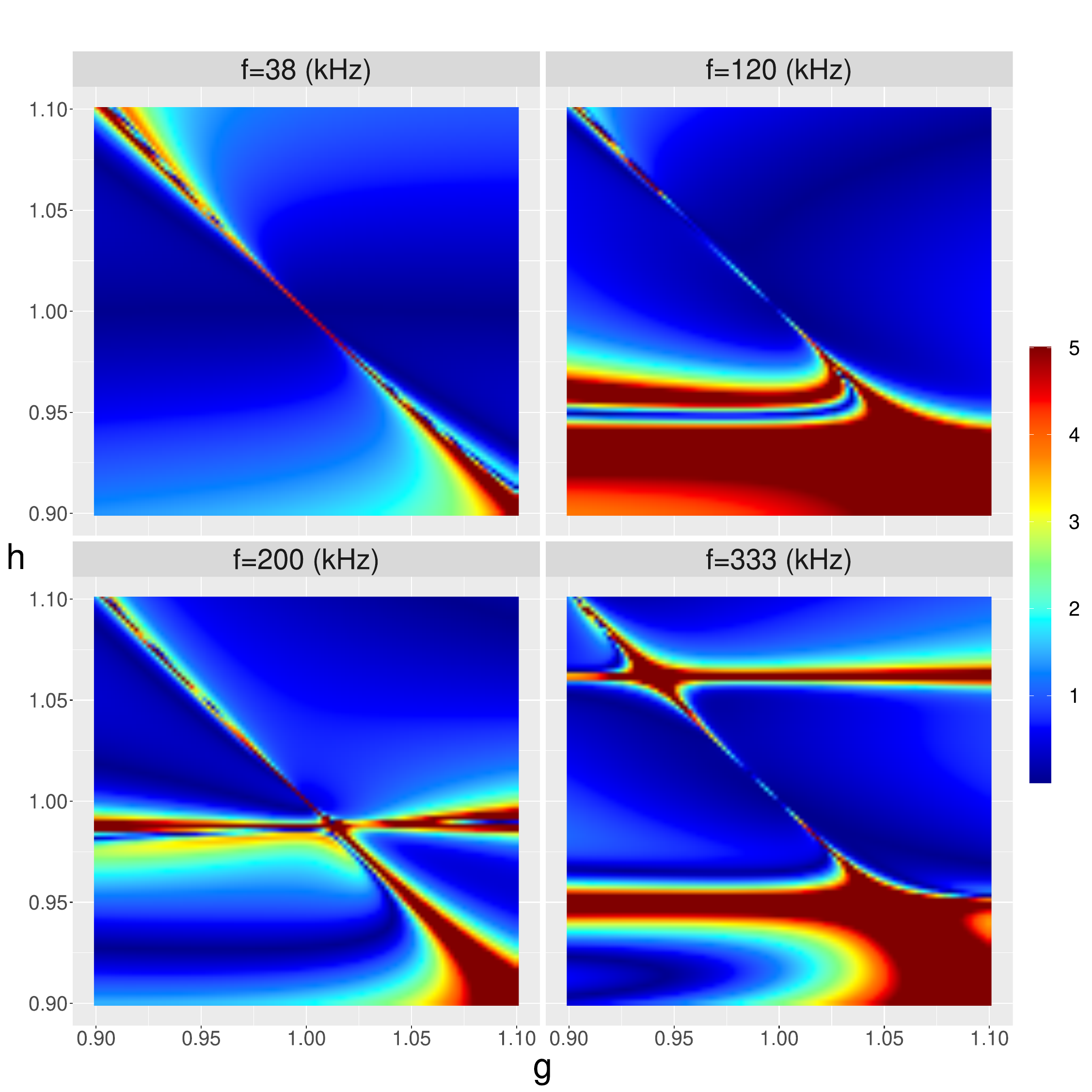}
	\caption{Absolute error $| \TS(f_\infty) -\TS(f_\infty^\text{ex})| $ for the backscattering with TetraScatt executed on a spherical mesh of $N= 9521$ elements with respect to the exact solution of a fluid sphere. Results for $f=$ 38, 120, 200 and 333 kHz are shown in the top left, top right, bottom left and bottom right panels, respectively. All plots were saturated at 5 dB to provide a comparison on equal footing.}
	\label{fig_born_exacta_esfera}
\end{figure}

Heatmaps of absolute error in $\TS$ according to Eq. \eqref{error_ts} are shown in the panels of Fig. \ref{fig_born_exacta_esfera} for the frequencies $f=$ 38, 120, 200 and 333 kHz in the range (0.9,1.1) for both $g$ and $h$. The TetraScatt predictions were computed with a $N=$ 9521 tetrahedra mesh.
The central point $(g,h)=(1,1)$ is always excluded since it represents a no-scattering condition.

Colours were saturated at $\varepsilon =$ 5 dB to clearly define zones of unacceptable error (more than 5 dB).
The pattern is clearly not trivial, gaining in complexity with increasing frequency and there are curves of high error related to nulls or resonances in the fluid sphere scattering.
In the context of the DWBA for homogeneous spheres peaks error appears over the curve $g=g(h)$ in the $(g,h)$ space domain, corresponding to regions where the pre-factor $(\gamma_\kappa h^2 - \gamma_\rho)$ vanishes. 

The translational symmetry in the direction of parameter $g$, present at $f = 38$ kHz is already broken at $f = 120$ kHz.

\subsection{Prolate spheroid in end-on incidence}

The previous comparison were made for the sphere which is an isotropic object with constant curvature. To quantify the curvature influence on the scattering by the BA, two prolate spheroids of different aspect-ratios $e$ under end-on incidence were considered.

\begin{figure}[!ht]
	\centering
	\includegraphics[scale=0.35]{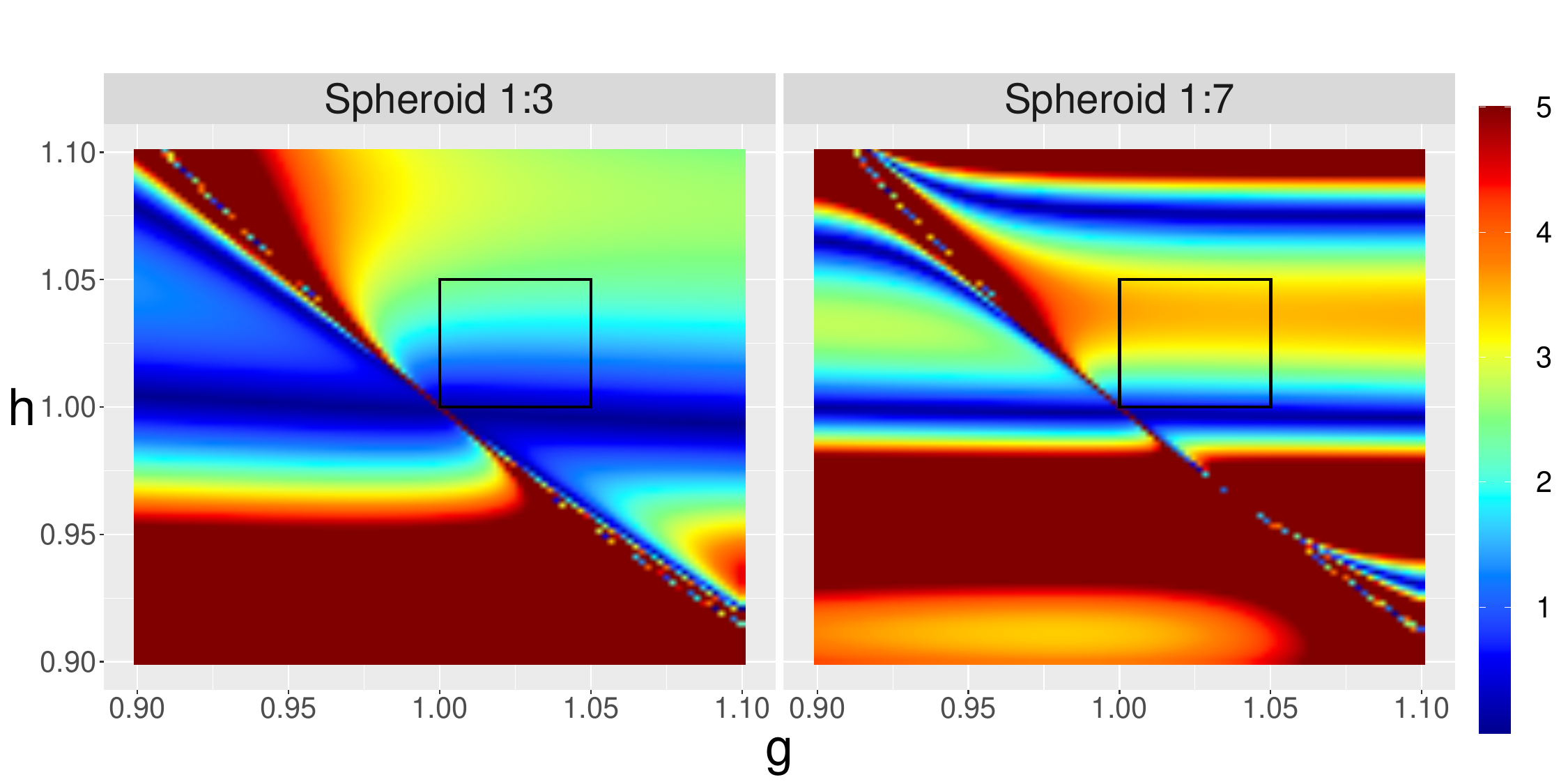}
	\caption{Absolute error $| \TS(f_\infty) -\TS(f_\infty^\text{ex})| $ for the end-on incidence backscattering at $f=$ 38 kHz with TetraScatt executed on a spheroidal mesh of $N=25574$ elements for 1:3 prolate spheroid (left panel) and 1:7 spheroid (right panel). All plots were saturated at 5 dB to provide a comparison on equal footing.
	}
	\label{fig_born_exacta_esferoide}
\end{figure}

Heatmaps of absolute error in $\TS$ for a prolate spheroid of 1:3 and 1:7 both evaluated at $f=$ 38 kHz are shown in the left and right panels of Fig. \ref{fig_born_exacta_esferoide}, respectively, in the range (0.9,1.1) for both $g$ and $h$. As observed in the previous subsection central point $(g,h)=(1,1)$ is excluded.
Spheroidal meshes of $N=$ 25574 for TetrasScatt evaluation of the 1:3 and 1:7 ratio spheroids were used for both cases.
Compared with the $f=$ 38 kHz sphere is noticeable as the pattern is more complex. The {\it low} error region is more restricted now, there are more regions where the saturated 5 dB error are reached.

As is known $g$ and $h$ should be close to $1$ to use Born approach but in practical applications it is worthwhile to investigate to what extents the approximation remains valid when these parameters deviates from the unitary value.
In the context of acoustic backscattering applied to aquatic ecosystem research, validity of these approximations for $g,h$ has been previously discussed. References \citen{chu1999phase,jones2009use,jech_comparisons_2015} suggest that the approximations can be applied for values of both $g$ and $h$ ranging from $1$ to $1.06$, approximately. In Fig. \ref{fig_born_exacta_esferoide}, a solid black line square $[1,1.05]\times[1,1.05]$ is included for comparison purposes.  It can be observed that the error can reach values up to 4 dB there, indicating that this solution may introduce significant errors when applied to an elongated organism, as the represented by our prolate spheroid in the end-on incidence.

It can be observed (also in the plot for the sphere case in the Fig. \ref{fig_born_exacta_esfera}) that regardless of the scatterer geometry or frequency, a significant error in the diagonal $g + h \approx 2$ appears. The reason for this behavior is explained by the factor $(\gamma_\kappa h^2 - \gamma_\rho)$ in Eq. \eqref{dwba_finf}. When this quantity is equal to zero the DWBA predicts $|f_\infty| = 0$, independently of the volume integral value.
By considering $g = 1 + \epsilon_g$ and $h = 1 + \epsilon_h$ where $\epsilon_g,\epsilon_h \ll 1$ and neglecting second-order terms in the Taylor expansion about $g=1, g=1$ is easy to show that
\[
	\gamma_\kappa h^2 - \gamma_\rho \approx -\frac{2}{g} \: ( g + h - 2 ),
\]
so that modeled $f_\infty$ and $\TS$ values are poorly estimated in the region $2/g \:( g + h - 2 )\approx 0$, specially when relative error is considered.
It is important to highlight the difference between the four quadrants in the $g,h$ space: it is fortunate that most of the modelling in the aquatic ecosystem research correspond to the first quadrant, i.e., parameters $g,h > 1$, where the Born approach for spheres and spheroids at least shows a better behaviour regarding absolute error.

\section{Application to Argentinian native species: peisos}
\label{peisos}

\begin{figure}[!ht]
 	\centering
 	\includegraphics[scale=0.2]{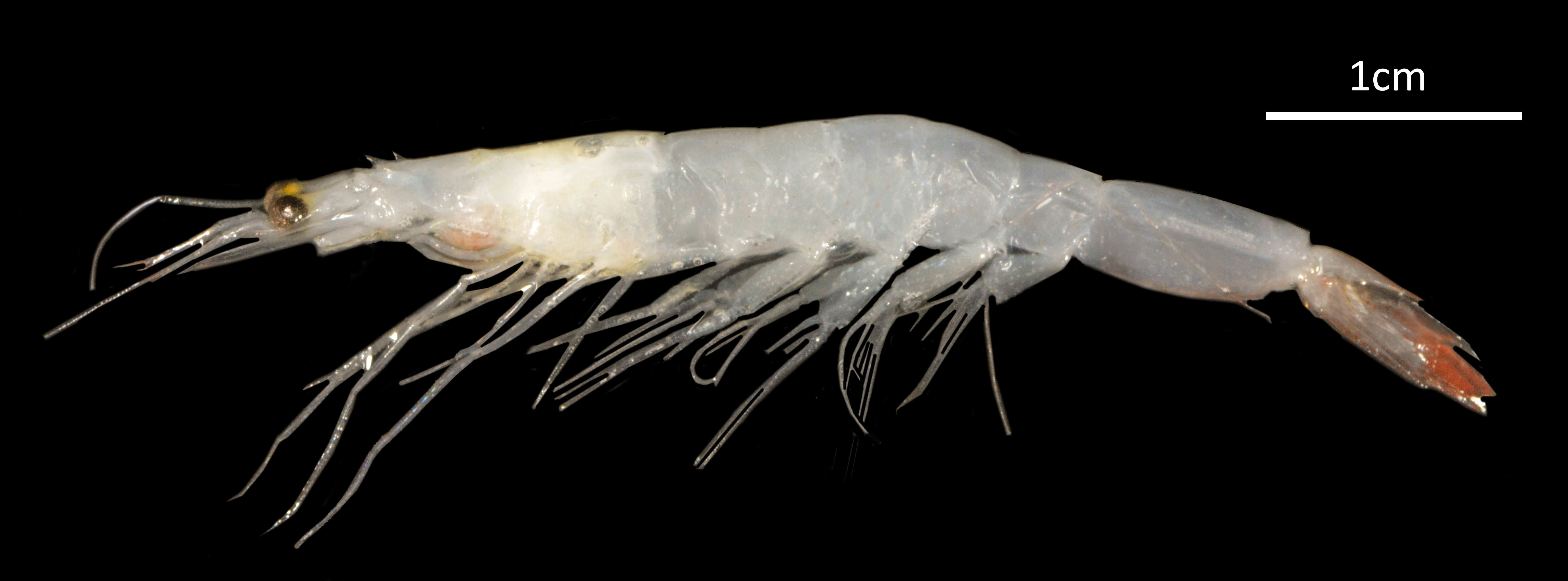}
 	\includegraphics[scale=0.2]{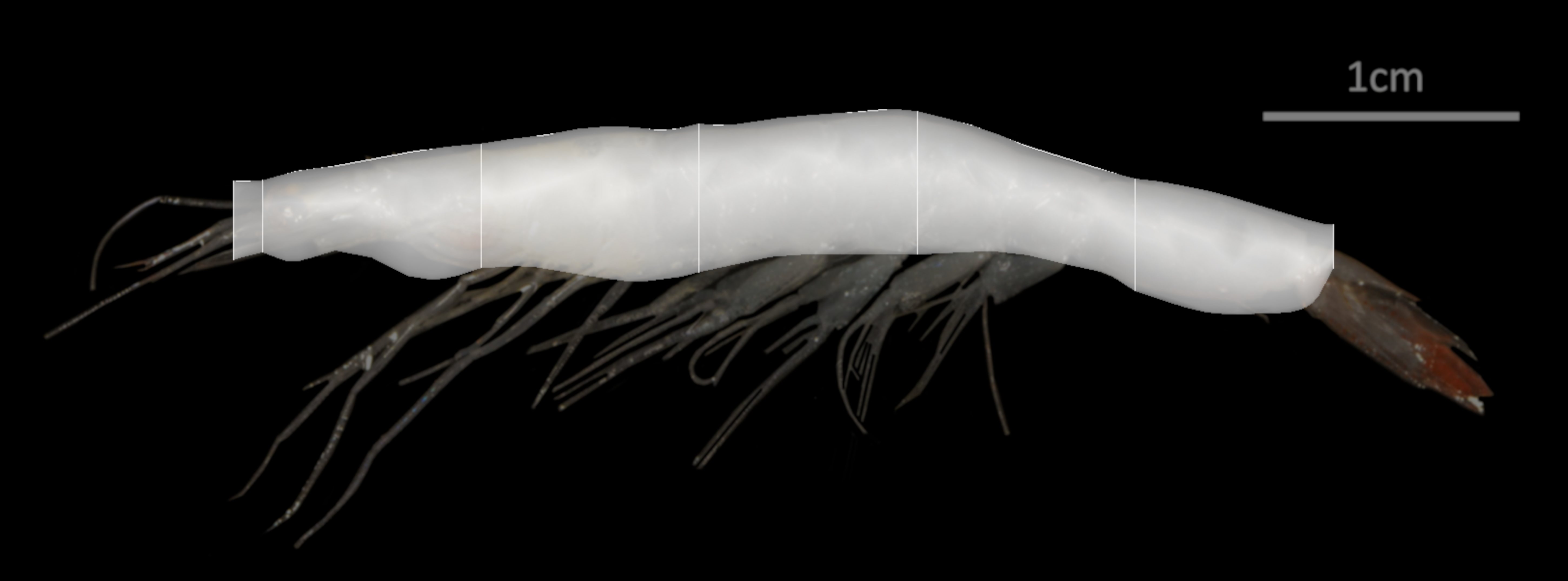}
 	\caption{Typical Peisos profile (top) and representative mesh with simplified surfaces and legs and tail removed (bottom).}
	\label{fig_peiso_lateral}
\end{figure}

The sergestid {\it Peisos petrunkevitchi} is a crustacean with abundance on the Argentinean sea of the Patagonia. There are part of the diet of the hake and other fish species.
The Fig. \ref{fig_peiso_lateral} shows a typical profile of an adult specimen (top view). Assuming that physical properties of Peisos are in the validity region of the Born approximation a representative volumetric mesh was built (see bottom view of Fig. \ref{fig_peiso_lateral}) and the backscattering according to TetraScatt and BEM were calculated.

Both the TetraScatt and BEM methods requires meshes of the scatterer's geometry, but their idiosyncrasies differ significantly.
In TetraScatt, the mesh is volumetric, and the solution is calculated  through simple integral evaluations. On the other hand, the BEM method employs a surface mesh representing the body, with the solution calculated from a matrix system solving procedure. The elements of this matrix are the outcome of integrals computed using quadrature methods.
BEM scattering calculations are significantly more time-consuming and require greater computational resources. However, the advantage lies in its ability to handle various contrast values, unlike the Born approximation.

\begin{figure}[!ht]
 	\centering
	\includegraphics[scale=0.201]{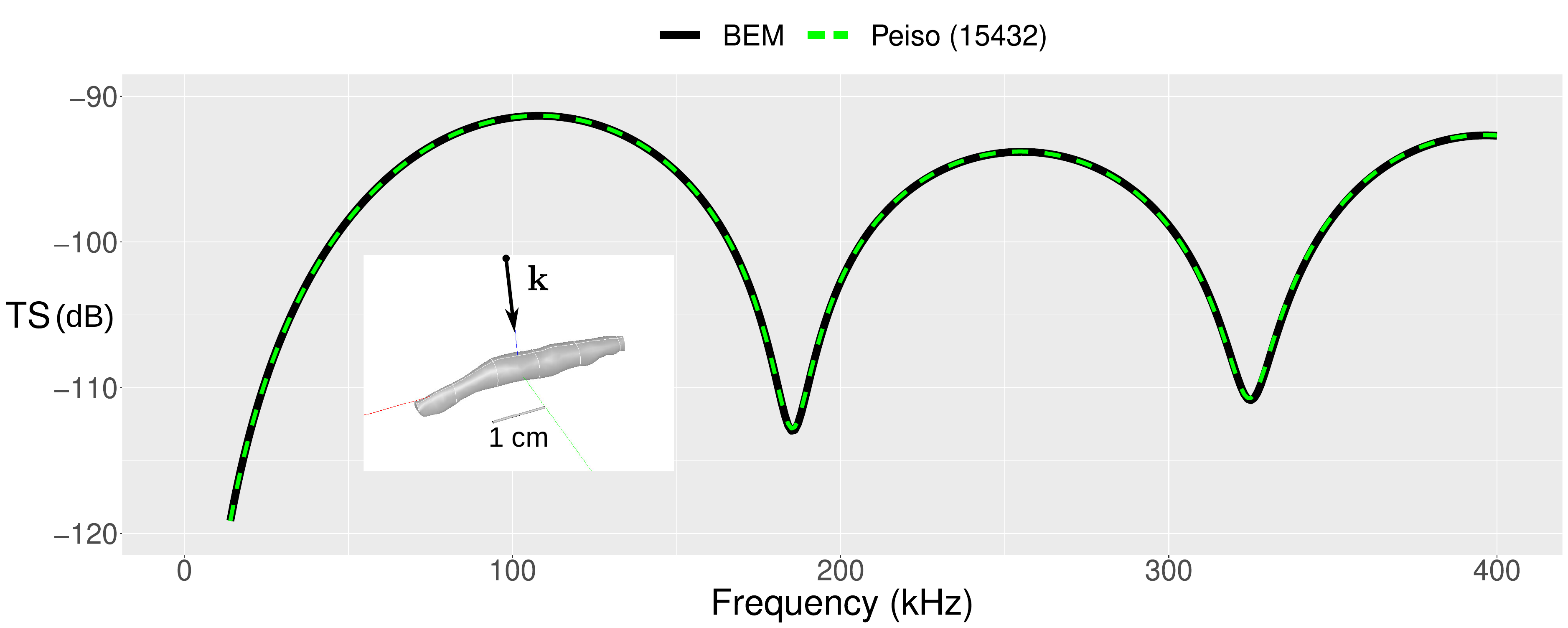}
	\caption{TS value for the backscattering by the mesh of figure \ref{fig_peiso_lateral} (bottom) representing a typical Peiso organism. The incidence is illustrated within the insert on the figure.}
	\label{fig_ts_peisos}
\end{figure}

Since the $\rho,c$ values appropriate for Peisos are currently not known, the calculations were carried out by using the Table 1 values.
The backscattering TS vs frequency pattern is shown in Fig. \ref{fig_ts_peisos}, where a good agreement between TetraScatt with a mesh of $16938$ tetrahedral elements and BEM with a mesh of $5306$ planar triangular elements is achieved.

For a quick comparison in computation resources used by BEM and TetraScatt it is interesting comment about the computation times. In this case, for the frequency interval considered in the Figure \ref{fig_ts_peisos}, which correspond to 389 values, TetraScatt took 3 seconds running in a single core of an Intel i7-12700 whereas BEM took one hour for the same task in a parallel execution on a 130 cores mixed cluster environment (Xeon E5-2620 and Xeon Gold 6240R).

\section{Conclusions}
\label{conclusions}

A model based on the DWBA that describes the backscattering of weakly scattering objects was developed,
computationally implemented in the R language and uploaded to a public repository.
The presented TetraScatt model has been validated against benchmark solutions based on analytical
integration and on general scattering methods (BEM and a previously developed scattering spheroidal code
based on spheroidal wave function calculation, which are not restricted to a weakly condition).
A study on the range of validity of the DWBA with respect to the contrast parameters $g, h$ in terms of
an error in dB showed a complicated pattern which hampers to obtain an easy prescription for the safe
application of the model regarding an error threshold.

Since the integration is exact in each tetrahedron any error in the scattering response (assuming that the weakly condition is fulfilled) must come from the inaccurate representation of the surface curvature of the bodies whose scattering is being calculated.
In any case larger volumetric meshes required to overcome this curvature issue doesn't handicap the method noticeably because the computational task to be done (integration reduces to a formula) is elementary and its complexity doesn't rise with an increase of frequency.
Instead, BEM formulation requires the solving of a matrix system whose size is related to the number of elements of the mesh and the elements themselves are evaluated through integration by quadrature rules which are increasingly expensive with the frequency.
Off course, BEM is an exact method in the sense that no approximations or limitations are assumed in the scattering problem besides a proper relationship between the triangle's segment length and the wavelength of the incident field.

The  TetraScatt only requires a volumetric mesh in the Gamma Mesh Format (GMF) to be able to compute the scattering without any kind of limitations about the geometrical shape of the object.
This last feature represent an add to the tools which are available to the underwater acoustic community and it is based on remarkably simple ideas on modelling and requires only conventional computational resources to be executed.

\clearpage





\appendix

\section{Function $F$}
\label{app_F}

Considering the following notational conventions
\begin{align*}
	kab = k_1\hat{\bf{k}}\cdot\vb{AB}, \quad kac = k_1\hat{\bf{k}}\cdot\vb{AC}, \quad kad = k_1\hat{\bf{k}}\cdot\vb{AD} \\
	kbc = k_1\hat{\bf{k}}\cdot\vb{BC}, \quad kbd = k_1\hat{\bf{k}}\cdot\vb{BD}, \quad kdc = k_1\hat{\bf{k}}\cdot\vb{DC},
\end{align*}
\begin{align*}
	ka = k_1\hat{\bf{k}}\cdot\vb{A}, \quad kb = k_1\hat{\bf{k}}\cdot\vb{B},
	\quad kc = k_1\hat{\bf{k}}\cdot\vb{C}, \quad kd = k_1\hat{\bf{k}}\cdot\vb{D},
\end{align*}
where it is supposed that $kad \neq 0$, (otherwise it is always  possible rearrange the four vertices in a way that this condition is fullfiled), then the function $F(\vb{A},\vb{B},\vb{C}, \vb{D},\hat{\bf{k}},k_1)$ which is results from the integration prescribed by the Eq. \eqref{integral_born_app} is given by
\[
F = \begin{cases}
	    \displaystyle \: -\frac{1}{4 kab} \left[
	     \frac{ e^{i(kd+kc)} \sinc(kdc) - e^{i(kb+kc)} \sinc(kbc) }{kbd} \: + \right. \\
			\\
		\displaystyle \hspace{14em}	\left.  \frac{e^{i(ka+kc)} \sinc(kac) - e^{i(kd+kc)} \sinc(kdc) }{kad}
	    \right], \qquad |kbd| \geq \varepsilon \\
	    \\
		\displaystyle \: -\frac{1}{4 kab} \left[
			e^{i(kb+kc)} \: \Gamma( kbd, kbc ) \: + \: \frac{ e^{i(ka+kc)} \sinc(kac) - e^{i(kd+kc)} \sinc(kdc)}{kad}
		\right], \qquad |kbd| < \varepsilon \\
    \end{cases}
\]
where $\varepsilon = 1$E-8 is a fixed parameter and the $\Gamma( x, y )$ function is defined according to
\[
	\Gamma( x, y ) = \frac{ e^{ix} \sinc(y-x) - \sinc(y) }{x}.
\]
To ensure proper behaviour when $x,y \leq \varepsilon$ appropriate Taylor expansions are used in the numerical implementation.
The {\it sinc} function is defined as $\sinc(x) = \sin(x)/x $.

\end{document}